\documentclass[doublecol]{epl2}

\title{A test-tube model for rainfall}

\author{Michael Wilkinson }

\institute{
  {Department of Mathematics and Statistics,
The Open University, Walton Hall, Milton Keynes, MK7 6AA, England}
}

\abstract{
If the temperature of a cell containing two partially miscible
liquids is changed very slowly, so that the miscibility is decreased, microscopic
droplets nucleate, grow and migrate to the interface due to
their  buoyancy. The system may show an approximately
periodic variation of the turbidity of the mixture, as the mean droplet size
fluctuates. These precipitation events are analogous to rainfall from warm clouds.
This paper considers a theoretical model for these experiments.
After nucleation the initial growth is by Ostwald ripening, followed by
a finite-time runaway growth of droplet sizes due to larger droplets
sweeping up smaller ones. The model predicts that the period
$\Delta t$ and the temperature sweep rate $\xi$ are related by
$\Delta t\sim C \xi^{-3/7}$, and is in good agreement with experiments.
The coefficient $C$ has a power-law divergence approaching the
critical point of the miscibility transition: $C\sim (T-T_{\rm c})^{-\eta}$, and
the critical exponent $\eta$ is determined.
}

\begin{document}

\maketitle

\section{Introduction}

Rainfall is a cyclic process, in which water vapour builds up in
the atmosphere, reaches supersaturation, and then forms droplets
of visible moisture which appear as clouds. The microscopic droplets
can grow by collisions and coalescence
until they are large enough to fall as rain, which removes the moisture from
the atmosphere.  The same mechanisms
will be applicable on other planets, but the role of water could be
taken by other small molecules. Many different processes may be involved \cite{Pru+97,Sha03}, and a quantitative understanding of the rainfall cycle has been lacking.
 In terrestrial rainfall, the growth of droplets through the size range $15-50\,\mu{\rm m}$ is
considered to be a \lq bottleneck' in the kinetics of rain initiation \cite{Sha03},
for which the mechanism is not yet fully understood.
It is valuable to investigate simplified models which can
serve as a benchmark against which complex phenomena
can be compared. As well as being important for terrestrial meteorology,
such models will be particularly important
in analysing weather phenomena on other planets, where the composition
and physical parameters of the atmosphere are relatively uncertain \cite{Sea+10,Eva+13}.

This paper discusses a laboratory model which includes the essential
features of the rain cycle. Two partially miscible liquids are placed
in a small cell, where they form two layers. The cell is placed in a computer
controlled thermostat, and its temperature is varied smoothly
away from the phase coalescence temperature $T_{\rm c}$, at a rate
described by a variable $\xi$ which has dimensions of inverse time
(defined by equation (\ref{eq: 4}) below). Several examples of this type of
experiment have been reported: \cite{Vol+97,Vol+99,Vol+02,Vol+07,Hay+08}.
The experiments show similar quantitative results: both layers show a variation
in turbidity, which is approximately periodic, with period $\Delta t$.
Periods of high turbidity end with microscopic droplets of fluid floating up or
down towards the interface. In general the upper and lower layers have
different periods, $\Delta t_{\rm t}$ and $\Delta t_{\rm b}$ respectively.
These periodic precipitation events are analogous to rainfall.

The experimental results reported in these earlier works do not span
a sufficiently wide range of parameters to definitively test quantitative predictions
about how the period $\Delta t$ depends upon the heating rate $\xi$.
However, a recent series of experiments reported in \cite{Lap11}
(see also \cite{Lap+11})
explores a very wide range of values of the experimental parameters,
and is consistent with 
\begin{equation}
\label{eq: 1}
\Delta t=C\, \xi^{3/7}
\end{equation}
where the multiplier $C$ is a function of the mean temperature $T$ during
the period, and is predicted to become independent of the height $h$ of the layer
when this is larger than some characteristic scale, $h_0$.
There is a slowing of the dynamics at the critical point,
with $C$ having a power-law divergence approaching
the miscibility transition at temperature $T_{\rm c}$: the data
in \cite{Lap11} are consistent with
\begin{equation}
\label{eq: 2}
C\sim (T-T_{\rm c})^{-\eta}\ ,\ \ \ \eta \approx 0.50
\ .
\end{equation}
Here I present the theoretical model leading to equation (\ref{eq: 1}), 
and discuss how $C$ is determined from the thermodynamic and 
kinetic coefficients of the mixture. The value of the
exponent $\eta$ in equation (\ref{eq: 2}) is determined from standard critical exponents.

The theory uses the Lifshitz-Slezov \cite{Lif+61,Lif+81} theory of Ostwald ripening
to describe the initial growth of droplets after they are nucleated.
When the droplets reach a certain size, their motion due to the
difference in density from the surrounding fluid becomes significant.
Buoyancy drives droplets towards the interface, with larger droplets
catching up and coalescing with smaller drops.
As droplets grow, their velocity towards the interface increases,
which further increases the rate at which they sweep up smaller droplets.
The model predicts a divergence of the droplet size in finite time, similar
to that which occurs in the evolution of rain droplets \cite{Kos+05}.

\section{The experimental system}

The system is a small transparent cell
containing two liquids (water and isobutoxyethanol were used in \cite{Lap11,Lap+11}), which are only partially miscible above a critical
temperature $T_{\rm c}$.
The phase equilibrium line for the two phases is a line in the
$\Phi$-$T$ plane, where $0\le \Phi\le 1$ is the volume-fraction of the denser
component. Above $T_{\rm c}$ the coexistence  curve has two branches,
denoted $\Phi_{\rm b}(T)$ and $\Phi_{\rm t}(T)$ for the bottom (denser) and top phases respectively.

As the system is heated, the miscibility decreases, implying that the minority component
will come out of solution in each phase. If the temperature change is extremely slow it may
diffuse into the lower layer, but as the system size is increased the material must precipitate out,
no matter how slow the temperature change. It is assumed that, immediately after a precipitation event,
the two layers are in equilibrium at points on the phase diagram, with compositions
$\Phi_{\rm b}$ and $\Phi_{\rm t}$ and with volumes $V_{\rm b}$ and $V_{\rm t}$.
A temperature
difference $\delta T$ causes a change in the equilibrium composition by
$\delta \Phi_{\rm b}<0$ and $\delta \Phi_{\rm t}>0$. Because the change is too rapid to
allow significant diffusion across the interface, the change in the volume of the lighter
component in the lower layer is $\delta v_{\rm b}=0$. This change has contributions
from the amount of the lighter component in solution, and from the formation of droplets
of total volume $\delta V_{\rm d}$, so that
\begin{equation}
\label{eq: 3}
\delta v_{\rm b}=0=(V_{\rm b}-\delta V_{\rm d})\Phi_{\rm b}+\delta V_{\rm d}\Phi_{\rm t}+
V_{\rm b}\delta \Phi_{\rm b}-V_{\rm b}\Phi_{\rm b}
\ .
\end{equation}
The most convenient measure of the rate of temperature change
is the rate of change of the volume fraction occupied by the droplets.
For the lower phase, this is
\begin{equation}
\label{eq: 4}
\xi_{\rm b}\equiv \frac{1}{V_{\rm b}}\frac{\delta V_{\rm d}}{\delta T}\frac{{\rm d}T}{{\rm d}t}
=\frac{1}{\Phi_{\rm b}-\Phi_{\rm t}}\frac{{\rm d}\Phi_{\rm b}}{{\rm d}T}\frac{{\rm d}T}{{\rm d}t}
\ .
\end{equation}

\section{A model}

Now consider a model for the cyclic fluctuations of turbidity.
Assume that at time $t_0$ a \lq rain' event
has occured and that this leaves a solution which is well-mixed and very close equilibrium, with
negligible supersaturation and with no visible particles in suspension. It is assumed
that there are always sufficient nucleation centres that
material can come out of equilibrium as the temperature increases
and reduce the supersaturation of the solution so that it is
always small. The resulting droplets of denser fluid will cause turbidity of the suspension if they
are sufficiently large, but while they are small compared
to the wavelength of light they might not be evident
in optical observations. The sub-microscopic droplets will grow as the apparent
supersatuation increases, so that at $t=t_0+\delta t$ there is an excess volume-fraction
$\xi \delta t $ in the form of sub-microscopic droplets.

Consider how to model the evolution of the radius of a droplet, $a(t)$.
After the droplets are nucleated they grow by diffusion of supersaturated
material onto their surface. As the droplets grow, the level of supersaturation
diminishes and the smallest droplets become unstable, because surface tension
increases their Laplace pressure. They are re-absorbed
into the solution and the material they contained diffuses onto the surface
of the larger droplets. This process is an example of \lq Ostwald ripening',
and it was first given a satisfactory treatment by Lifshitz and Slezov \cite{Lif+61}. Their
account treats a quenched solution, in which the total amount of excess material
in the supersaturated solution (having a volume fraction $V_{\rm d}$) is a constant. In our
problem the amount of excess material increases linearly with time: $V_{\rm d}=\xi (t-t_0)$.
This will result in a minor modification to the Lifshitz-Slezov theory when it is applied
to the experiment.

The essential features of the Lifshitz-Slezov approach can be summarised
as follows. For the sake of simplicity, a dilute solution approximation
is used, which is analytically tractable. The interior of a
droplet of radius $a$ has a pressure which is
higher than the ambient pressure by $\Delta p=2\sigma /a$, where $\sigma $
is the surface tension at the phase boundary. This increased pressure implies that
the minority component must have a higher concentration in order to be in
equilibrium: the chemical potential must be increased by $\Delta \mu=\Delta p v$,
where $v$ is the molecular volume of the minority component. Using a dilute solution
approximation, this increase in chemical potential is created by an increase in the volume
fraction, $\Delta \Phi$, satisfying $kT \Delta \Phi/\Phi_{\rm e}=\Delta \mu$, where $\Phi_{\rm e}$
is the equilibrium concentration. The surface of a droplet of radius $a$ is therefore in contact
with a layer of solvent with a volume fraction
\begin{equation}
\label{eq: 5}
\Phi(a)=\Phi_{\rm e}+\frac{\Lambda}{a}
\end{equation}
where $\Lambda$ is a Kelvin length, defined by:
\begin{equation}
\label{eq: 6}
\Lambda=\frac{2\sigma v\Phi_{\rm e}}{kT}=\frac{2\sigma V_{\rm m}}{RT}\Phi_{\rm e}
\end{equation}
(here $V_{\rm m}$ is the molar volume of the minority component,
and $R$ is the universal gas constant).
The concentration in the vicinity of each droplet is in 
quasi-equilibrium and may be approximated
by a solution of Laplace's equation. The concentration
far from a droplet is $\Phi(\infty)=\Phi_{\rm e}+s$, where $s$ is the supersaturation of the
solution. The concentration at a distance $r$ from the nearest
droplet is approximated by
\begin{equation}
\label{eq: 7}
\Phi(r)=\Phi_{\rm e}+s+\frac{\Lambda-sa}{r}
\ .
\end{equation}
There is a diffusive flux of material onto the droplet surface, which
causes its radius to change at a rate
\begin{equation}
\label{eq: 8}
\frac{{\rm d}a}{{\rm d}t}=D\frac{\partial \Phi}{\partial r}\bigg\vert_{r=a}=
\frac{D\Lambda}{a}\left (\frac{1}{a_0}-\frac{1}{a}\right)
\end{equation}
with $a_0=\Lambda/s$, where $D$ is the interdiffusion
coefficient of the two components. The interpretation is that droplets smaller
then $a_0$ shrink under the effects of the Laplace pressure, and those
larger than $a_0$ grow by absorption of material evaporating
from the smaller droplets. The supersaturation must decrease
as a function of time so that the largest droplets can continue to
grow. It follows that $a_0$ is comparable to the typical
droplet size (and in fact the Lifshitz-Slezov theory predicts that
$\langle a\rangle =a_0$ \cite{Lif+61}). For the growth of the
largest droplets, the $a_0$ term in (\ref{eq: 8})
can be neglected, implying that $a^3\sim D\Lambda t$. In fact the Lifshitz-Slezov
theory predicts that the mean droplet size is \cite{Lif+61}
\begin{equation}
\label{eq: 9}
\langle a(t)\rangle = \left[\frac{4}{9}D\Lambda (t-t_0)\right]^{1/3}
\ .
\end{equation}
Here the dynamics is a little different from the standard
Ostwald ripening process, where the system is quenched at a fixed
temperature. In that case the amount of material which comes
out of solution is independent of time, and the number density of droplets
decreases due to evaporation as the mean droplet size increases.
In the case treated here, however, the amount of material which comes out of solution
is proportional to time. Together with a growth law of the form $a\sim t^{1/3}$,
this implies that the droplet number remains constant in the long-time limit. The droplet
density is set by competitative growth in the early stages, but at large times
droplets do not evaporate and their number density approaches a constant.

The diffusive growth of droplets continues until other effects
become significant, creating macroscopic motion. The first such effect to become relevant
is when droplets start to settle towards the interface due to their density contrast.
The sinking velocity in the upper layer, $u_{\rm t}$, may be estimated by balancing
gravitational forces and viscous drag. Using the Stokes formula for the drag on a small
sphere gives 
$\frac{4\pi}{3}(\rho_{\rm b}-\rho_{\rm t})ga^3=6\pi \rho_{\rm t}\nu_{\rm t} a u_{\rm t}$ 
where $\nu_{\rm t}$ is the kinematic viscosity of the lighter fluid, and
where the densities of the lighter and heavier phases are $\rho_{\rm t}$ and $\rho_{\rm b}$ respectively. In the following $\Delta \rho=\rho_{\rm b}-\rho_{\rm t}$.
The settling speed is
\begin{equation}
\label{eq: 10}
u_{\rm t}=\frac{2}{9}\frac{\Delta \rho g}{\rho_{\rm t}\nu_{\rm t}}a^2\equiv \kappa_{\rm t} a^2
\ .
\end{equation}
The gravitational settling allows the rate of accumulation of material on a droplet
to increase because the larger droplets, which fall at a faster rate, can overtake the smaller droplets.
 If droplets collide they can merge, and the merged droplet will fall at an even larger rate.
This sweeping process will be modelled by assuming that a droplet has already
grown to a size $a(t)$ where it is much larger than the other droplets in its
path. The settling velocity of the smaller droplets may be neglected. The larger
drop moves through a \lq gas' of smaller droplets which occupy a volume fraction
$\xi (t-t_0)$, and they cause the lower surface to sweep up a volume per unit time
equal to $\dot v_{\rm d}=\pi a^2 u \epsilon\xi (t-t_0)$, where $\epsilon$ is a collision
efficiency. This rate of growth of the volume implies a
rate of growth of the droplet radius given by $\dot v_{\rm d}=4\pi a^2 \dot a$, so that
the droplet radius grows by particle accretion at a rate $\dot a=\xi t \epsilon u/4$, where
$u$ is the settling speed (given by (\ref{eq: 10}) for the droplets sinking in the upper layer).

Now consider a model for the time evolution of the radius of the
largest droplets. Noting that $a_0\sim \langle a\rangle$, growth of the largest droplets may be described
by the relation $\dot a\sim D\Lambda/a^2$. Combining this with the relation
for growth by sweeping gives the model equation
\begin{equation}
\label{eq: 11}
\frac{{\rm d}a}{{\rm d}t}=\frac{D\Lambda}{a^2}+\frac{\epsilon}{4}\kappa a^2\xi t
\ .
\end{equation}
This equation suggests that the
evolution of the droplets can be divided into two stages, depending upon which
of the two terms in the expression for $\dot a$ is dominant. The first stage, involving
growth by Ostwald ripening and diffusional accretion, lasts for a time $t_1$, which is determined
by the condition that the two terms in (\ref{eq: 11}) become equal. This condition
is satisfied when the droplets reach a size $a_1$. From this point on it will be assumed
that the collision efficiency is of order unity, and other dimensionless constants will
be dropped. The condition for the crossover is
\begin{equation}
\label{eq: 12}
\frac{D\Lambda}{\kappa \xi}\sim  a_1^4 t_1
\ .
\end{equation}
However, from (\ref{eq: 9}) it is found that $a_1^3\sim D\Lambda t_1$, so that
solving for $t_1$ gives $\frac{D\Lambda}{\kappa\xi}\sim \left(D\Lambda t_1\right)^{4/3}t_1$,
and hence
\begin{equation}
\label{eq: 13}
t_1\sim \left(\frac{1}{D\Lambda (\kappa\xi)^3}\right)^{1/7}
\ .
\end{equation}
In the second stage, the equation (\ref{eq: 11}) for droplet growth may be approximated
by $\dot a/a^2=\kappa \xi (t-t_0)$. This has a solution (with initial condition $a=a_1$
at $t=t_1+t_0$)
\begin{equation}
\label{eq: 14}
\frac{1}{a_1}-\frac{1}{a}=\frac{\kappa \xi}{2}[(t-t_0)^2-t_1^2]
\ .
\end{equation}
According to this solution, $a(t)$ diverges in a finite time, so that
$a\to \infty$ as $t\to t_1+t_2$, with $t_2$ having the same scaling as $t_1$.
In summary, the model predicts that the period $\Delta t=t_1+t_2$ is
\begin{equation}
\label{eq: 15}
\Delta t \approx \alpha \left(\frac{1}{D\Lambda\kappa^3}\right)^{1/7}\xi^{-3/7}
\end{equation}
where $\alpha$ is a dimensionless prefactor. Because of the
finite-time divergence of the droplet sizes, the timescale for the
oscillations is not predicted to depend upon the size of the cell.
But the cell does need to be large eonugh for the finite-time runaway
growth to happen.

Equation (\ref{eq: 15}) was compared with experimental data
in the Ph.D. thesis of T. Lapp \cite{Lap11}, where it was shown to
give a very good description of the experimental data
for water/isobutoxyethanol. The rate of change of order parameter $\xi$ was
varied over nearly three decades and the temperature relative to
the critical point, $T-T_{\rm c}$,  by nearly three decades.
Another observation which supports the theory is that the periods
of high and low turbidity are of a comparable extent, which is consistent
with the theoretical prediction that $t_1\sim t_2$.

The experimental data in \cite{Lap11} show that the period $\Delta t$ diverges as
we approach the critical temperature, $T_{\rm c}$, in accordance
with equation (\ref{eq: 2}), with the exponent $\eta$ taking the
same value for both layers. This divergence is
a consequence of the fact that the coefficient $C$ contains three
factors which vanish at the critical point. The density
difference between the phases, $\Delta \rho$, has the
critical behaviour of an order parameter, with exponent $\beta$.
The critical exponent for the interfacial surface tension, $\sigma$,
can be related to the correlation length exponent $\nu$ \cite{Wid65}. The
van Hove theory of phase transitions indicates that the critical exponent of the interdiffusion
coefficient is the susceptibility exponent $\gamma$ \cite{Hoh+77}
(a different value has been proposed \cite{Bin+07}, but $\gamma$ appears closer to
experimental measurements \cite{Ste+95} on the system used in \cite{Lap11}).
 For a three-dimensional system with a one-component conserved order parameter, the exponents are:
\begin{eqnarray}
\label{eq: 16}
\Delta \rho \sim |T-T_{\rm c}|^\beta & \quad\quad & \beta \approx 0.327
\nonumber \\
\sigma \sim |T-T_{\rm c}|^{2\nu} & \quad \quad & \nu \approx 0.630
\nonumber \\
D\sim |T-T_{\rm c}|^\gamma  &\quad\quad & \gamma \approx 1.237
\ .
\end{eqnarray}
From equations (\ref{eq: 6}), (\ref{eq: 10}) and (\ref{eq: 15}), the
critical exponent for the coefficient $C$ is
\begin{equation}
\label{eq: 17}
\eta=\frac{3\beta+2\nu +\gamma}{7}\approx 0.498
\end{equation}
which is also in satisfactory agreement with the experimental results \cite{Lap11}.

\section{Discussion}

It has been argued that the periodic precipitation
phenomenon which has been described in several works
 \cite{Vol+97,Vol+99,Vol+02,Vol+07,Hay+08,Lap11,Lap+11} is analogous
to an atmospheric precipiation cycle in a stable atmosphere.
Consider whether the same mechanisms are relevant to the growth of 
real rain droplets. It has been argued \cite{Pru+97} that there is a
\lq bottleneck' in the growth kinetics of raindrops
at radii in the range $a\approx 15-50\,\mu{\rm m}$, between
smaller particle sizes where growth by condensation is
efficient, and larger particle sizes where collisions due
to gravitational settling become important. The Ostwald
ripening process may allow this gap to be bridged,
because it allows droplets to grow by diffusive
transfer of material from smaller droplets, without
the necessity for collisions. Note that the growth law, given by
equation (\ref{eq: 9}), is independent of the density of the
droplets. The physical parameters which are required
to estimate the growth law are, at $10^\circ {\rm C}$,
diffusion constant for water vapour in air,
$D=2.4\times 10^{-5}\,{\rm m}^2{\rm s}^{-1}$,
surface tension $\sigma=7.4\times 10^{-2}\,{\rm Nm}^{-1}$,
molar volume $V_{\rm m}=1.8\times 10^{-5}\,{\rm m}^3$,
saturation volume fraction $\Phi_0=1.2\times 10^{-5}$. These
give a Kelvin length $\Lambda=1.4\times 10^{-14}\,{\rm m}$
and consequently, using (\ref{eq: 9}), the growth law is
$a(t)\approx 6\times 10^{-9} (t/{\rm s})^{1/3}$. For water
droplets in the air, losing latent heat by conduction is another
significant effect, and $D$ must be replaced by an effective diffusion
coefficient $D_{\rm eff}<D$ \cite{Mas57}. These estimates imply
that growth to a radius of $25\,\mu {\rm m}$ by Ostwald ripening
requires several days, so that Ostwald ripening is too slow
to be important for terrestrial clouds. Clement \cite{Cle08}
reached the same conclusion about the relevance of
Ostwald ripening to terrestrial rainfall.

Finally, a speculative remark. Extra-solar planets are being discovered
at a prodigious rate \cite{But+06}, and that techniques are being developed to
identify their atmospheric composition \cite{Sea+10,Eva+13}. The sizes and orbital
parameters of these planets are highly variable, and it seems likely
that many of them will exhibit exotic weather phenomena.
On Earth,  rainfall is chaotic and unpredictable. Other planets, however,
may have temporally periodic rainfall events driven by Ostwald
ripening, analogous
to the test-tube model considered here.

\acknowledgments{ I have benefitted from discussions
with J. Vollmer, T. Lapp and M. Rohloff (MPI DS G\"ottingen), who
explained the derivation of their control parameter $\xi$, equation (\ref{eq: 4}),
and who tested (\ref{eq: 15}) against their experimental data. }

\end{document}